\documentstyle[jkas]{article}

\beginpage{169}
\endpage{182}
\year{2010}\volume{43}\month{October}

\runningauthor {S. KIM ET AL.}
\runningtitle{WIDE-FIELD OFF-AXIS TELESCOPE}

\month{October} \year{2010} \volume{43} \issueno{5}
\beginpage{169}\endpage{182}
\date{Received April 10, 2010; Revised August 02, 2010; Accepted September 26, 2010}

\begin{document}
\title{Proto-Model of an Infrared Wide-Field Off-Axis Telescope}
\author{Sanghyuk Kim$^{1,2}$, Soojong Pak$^{2}$, Seunghyuk Chang$^{3}$, Geon Hee Kim$^{1}$, Sun Choel Yang$^{1}$, Myung Sang Kim$^{1}$, Sungho Lee$^{4,5}$, and Hanshin Lee$^{5}$}
\address{$^1$ Korea Basic Science Institute, Daejeon 305-333, Korea
\\ {\it E-mail : ksh83@kbsi.re.kr}}
\address{$^2$ School of Space Research, Kyung Hee University, 1 Seocheon-dong, Giheung-gu, Yongin-si, Gyeonggi-do 446-701, Korea
\\ {\it E-mail : soojong@khu.ac.kr}}
\address{$^3$ Samsung Electronics, Suwon-si, Gyeonggi-do 443-370, Korea}
\address{$^4$ Korea Astronomy and Space Science Institute, Daejeon 305-348, Korea}
\address{$^5$ McDonald Observatory, Univ. of Texas at Austin, Austin, TX 78712, USA}
\offprints{S. Pak}
\abstract{We develop a proto-model of an off-axis reflective telescope for infrared wide-field observations based on the design of Schwarzschild-Chang type telescope. With only two mirrors, this design achieves an entrance pupil diameter of 50 mm and an effective focal length of 100 mm. We can apply this design to a mid-infrared telescope with a field of view of $8\arcdeg \times 8\arcdeg$. In spite of the substantial advantages of off-axis telescopes in the infrared compared to refractive or on-axis reflective telescopes, it is known to be difficult to align the mirrors in off-axis systems because of their asymmetric structures. Off-axis mirrors of our telescope are manufactured at the Korea Basic Science Institute (KBSI). We analyze the fabricated mirror surfaces by fitting polynomial functions to the measured data. We accomplish alignment of this two-mirror off-axis system using a ray tracing method. A simple imaging test is performed to compare a pinhole image with a simulated prediction.}

\keywords{Instrumentation: miscellaneous --- methods: laboratory --- techniques: miscellaneous --- telescopes --- infrared: general}
\maketitle


\section{INTRODUCTION}

Since the discovery by William Herschel in 1800, infrared (IR) light has been used extensively for commercial and military purposes as well as for astronomical observations. For IR telescopes and instruments, it is difficult in general to make an optical system using refractive lens components, because they can not avoid chromatic aberration and absorption bands over a wide IR wavelength range. Additionally, refractive systems are not appropriate for big telescopes since fabrication and mount of a large-sized lens is tricky \citep{lee92}. Reflective optics does not suffer from these problems. But conventional on-axis reflective telescopes have some intrinsic deficiencies in imaging performance. Secondary mirror obstruction causes scattering, diffraction and stray light \citep{chang03, chang06, moretto99, moretto04}. Scattering has two problems: (1) enhanced photon noise due to excessive scattered light and (2) time-variable systematic noise due to a complicated point spread function (PSF) of the telescope \citep{kuhn99, kuhn00}. Measurements over a very wide photometric dynamic range (e.g. astronomical observations of faint objects near bright sources such as extra-solar planet detection) are often limited by the characteristics of scattered light of a telescope \citep{moretto04}. The secondary obstruction also imposes restrictions on a wide field of view (FOV) which requires a large-sized secondary mirror \citep{moretto99}.

We are developing a wide-field IR telescope for observations of diffuse interstellar medium and cosmic background radiation in the mid-IR wavelength regime. To avoid the above shortcomings of refractive optics or on-axis reflective systems, we choose an off-axis reflective telescope as an alternative solution. As shown in the following section, this off-axis design uses a small given space efficiently and can be applied to a payload on micro satellites like STSAT series or sounding rockets \citep{bock06, lee05, lee06, yuk07}. Recently, several off-axis telescope projects have been initiated; e.g. the New Off-axis 1.7 m Solar Telescope \citep{didkovsky04} and the Advanced Technology Solar Telescope \citep{keil03}.

To properly design an off-axis telescope with a wide FOV, a careful analysis of aberration characteristics is needed. There have been extensive researches on non-symmetric optical systems \citep{sands72, sasian94, stone94, araki05, nakano05}. However, a closed form solutions for the mirror shapes and system configurations in order to maximize FOV of two-mirror off-axis optical systems became available only recently \citep{chang03, chang05, chang06, chang2006}. Chang et al. showed that the dominant aberration of a confocal off-axis reflective imaging system is linear astigmatism and derived its aberration coefficient in terms of mirror shape and system configuration parameters. They proved that an off-axis two-mirror telescope can have an equivalent aberration performance to a conventional on-axis telescope by designing the system to satisfy an equation derived for the elimination of linear astigmatism.

\begin{figure}[t]
\centering
 \epsfxsize=6.7cm
 \epsfbox{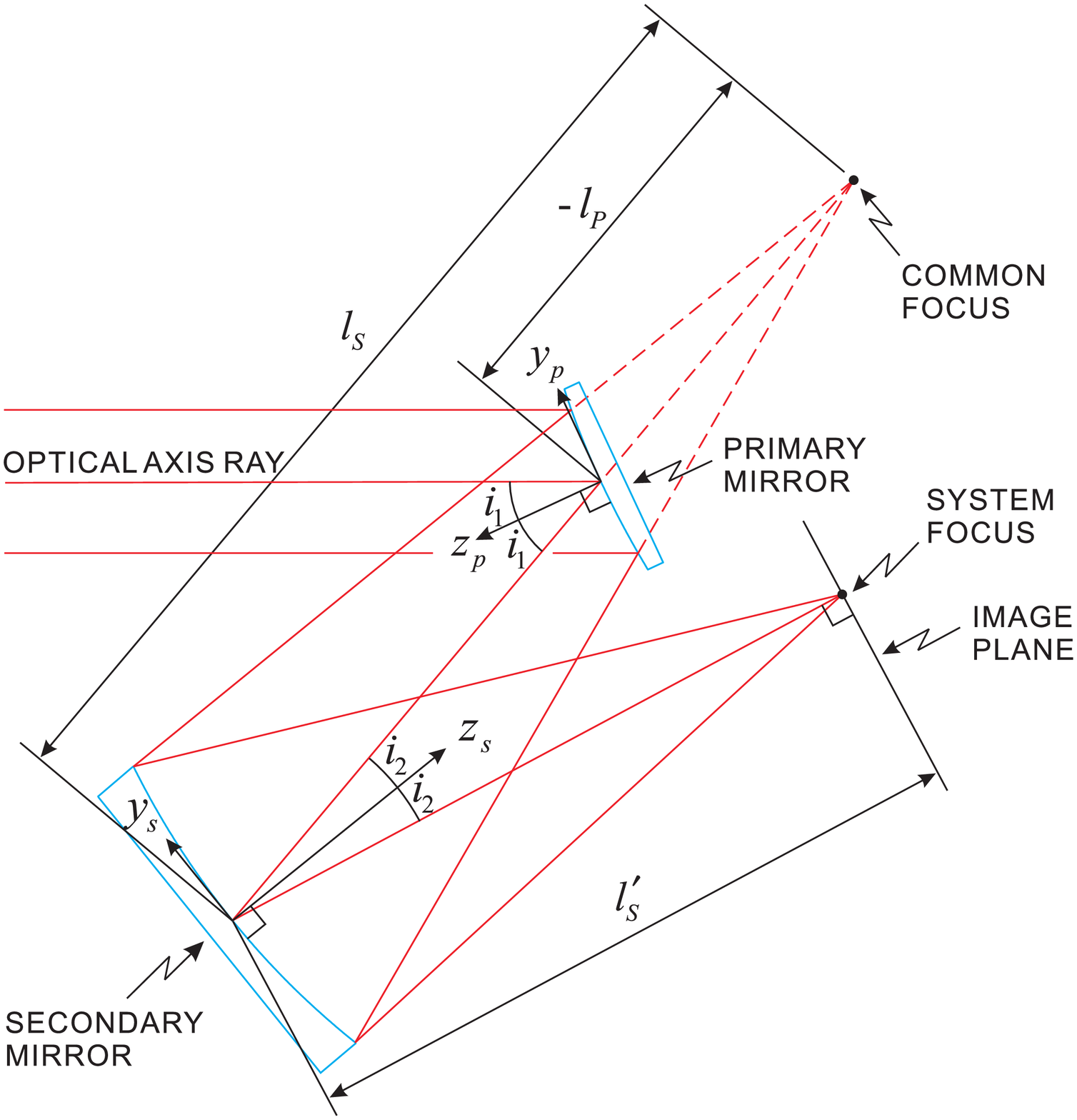}
\caption{Optical layout of the Schwarzschild-Chang off-axis telescope.}
\end{figure}

\begin{figure}[t]
\centering
 \epsfxsize=8cm
 \epsfbox{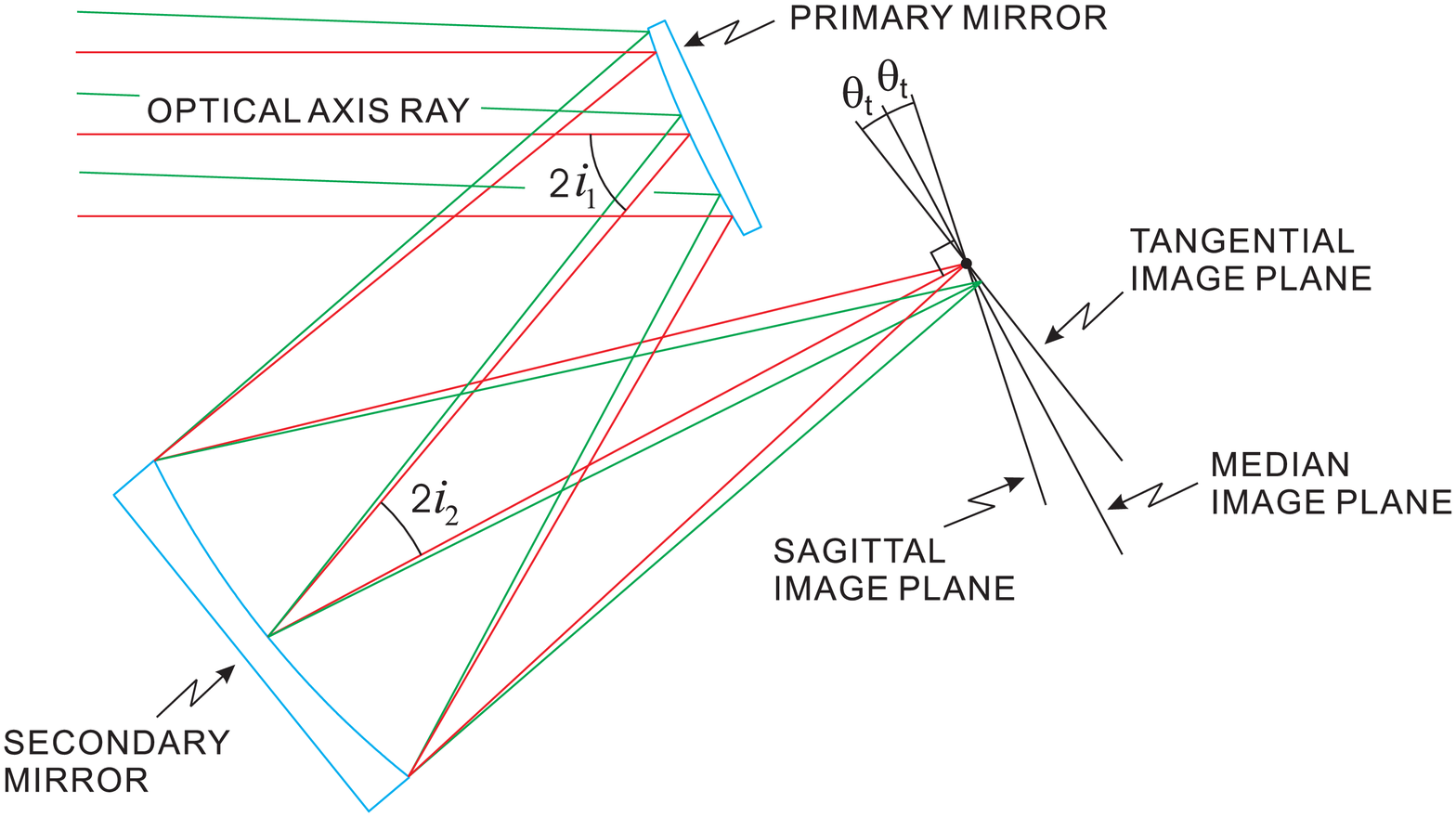}
\caption{Stop of the Schwarzschild-Chang off-axis telescope.}
\end{figure}

Off-axis aspheric mirrors can be manufactured in two different ways. The traditional method is to cut a small part of conic section from a large aspheric mirror. In this method we cannot make a mirror with a high offset value. On the other hand, an off-axis aspheric mirror can be directly manufactured as an off-axis segment. In this case, there are no restrictions on the offset and the central surface of the extracted mirror can be parallel to the back surface \citep{lee92}. The fabrication of non-axially symmetric off-axis mirrors is now possible by using the advances of diamond turning technology \citep{kim07, yang07}.

Another issue of an off-axis telescope is its optical alignment which has more degrees of freedom (DOF) than for an on-axis
telescope. Off-axis telescopes are usually aligned using interferometers or Hartmann tests. However, since alignment tolerance for an IR telescope can be much more relaxed than for an optical telescope, the interferometer method using laser beam may not be necessary for our IR system. Thus we develop an alignment system based on a simple ray tracing method.

In this paper, we present proto-model development of an IR wide-field off-axis telescope. In Section 2, we describe the optical design and specification of our Schwarzschild-Chang type telescope. Section 3 details the fabrication, measurement and analysis of the off-axis mirror surfaces. The alignment process is described in Section 4 and results from a point source simulation and test are presented in Section 5.

\section{SCHWARZSCHILD-CHANG OFF-AXIS \\ TELESCOPE}

The optical design of the manufactured telescope is a two-mirror off-axis wide-field system. It has a convex off-axis primary mirror and a concave off-axis secondary mirror. The design can be considered as an off-axis version of the on-axis Schwarzschild flat field anastigmat \citep{chang06, schroeder00}. We first design a classical off-axis inverse Cassegrain system \citep{chang05} such that its linear astigmatism is eliminated. Then, higher-order terms of the equations describing the mirror surfaces were optimized by an optical design software to yield maximum FOV performance.

\begin{table}[t]
\begin{center}
\centering
\caption{Main parameters of the Schwarzschild-Chang type wide-field off-axis telescope \label{tbl1}}
\doublerulesep2.0pt
\renewcommand\arraystretch{1.5}
\begin{tabular}{cc}
\hline \hline
   Parameter      &      Value\\ \hline
$l_{p}$  & -141.421 mm\\
$l_{s}$  & 341.421 mm\\
$l_{s}'$ & 241.421 mm\\
$i_{1}$  & 25 $\arcdeg$\\
$i_{2}$  & 10.9321 $\arcdeg$\\
Aperture Diameter & 50 mm\\
Focal Length & 100 mm\\ \hline
\end{tabular}
\end{center}
\end{table}

\begin{table}[p]
\begin{center}
\centering
\caption{Coefficients of the surface functions for off-axis mirrors \label{tbl2}}
\doublerulesep2.0pt
\renewcommand\arraystretch{1.5}
\begin{tabular}{crr}
\hline \hline
Coefficient$^{\rm a}$ & Primary & Secondary \\ \hline
$a_{2}	[mm^{-1}]$ & $1.950515\times10^{-3}$ & $1.800440\times10^{-3}$ \\
$a_{4}	[mm^{-1}]$ & $1.602141\times10^{-3}$ & $1.735687\times10^{-3}$ \\
$a_{6}	[mm^{-2}]$ & $2.914423\times10^{-6}$ & $-2.071217\times10^{-7}$ \\
$a_{8}	[mm^{-2}]$ & $2.393889\times10^{-6}$ & $-1.996724\times10^{-7}$ \\
$a_{9}	[mm^{-3}]$ & $4.435626\times10^{-8}$ & $6.827679\times10^{-9}$ \\
$a_{11} [mm^{-3}]$ & $7.571553\times10^{-8}$ & $1.315456\times10^{-8}$ \\
$a_{13} [mm^{-3}]$ & $3.262307\times10^{-8}$ & $6.345179\times10^{-9}$ \\
$a_{15} [mm^{-4}]$ & $1.452585\times10^{-10}$ & $-2.957232\times10^{-12}$ \\
$a_{17} [mm^{-4}]$ & $2.308069\times10^{-10}$ & $-5.813869\times10^{-12}$ \\
$a_{19} [mm^{-4}]$ & $1.153974\times10^{-10}$ & $-2.518645\times10^{-12}$ \\
$a_{20} [mm^{-5}]$ & $-2.427824\times10^{-12}$ & $4.291074\times10^{-14}$ \\
$a_{22} [mm^{-5}]$ & $-3.570415\times10^{-12}$ & $1.215566\times10^{-13}$ \\
$a_{24} [mm^{-5}]$ & $2.245922\times10^{-12}$ & $1.434325\times10^{-13}$ \\
$a_{26} [mm^{-5}]$ & $1.097917\times10^{-12}$ & $4.790632\times10^{-14}$ \\
$a_{28} [mm^{-6}]$ & $1.653002\times10^{-14}$ & $-2.659321\times10^{-18}$ \\
$a_{30} [mm^{-6}]$ & $1.361400\times10^{-13}$ & $1.757693\times10^{-16}$ \\
$a_{32} [mm^{-6}]$ & $9.315354\times10^{-14}$ & $8.534014\times10^{-17}$ \\
$a_{34} [mm^{-6}]$ & $6.917890\times10^{-15}$ & $-3.068882\times10^{-17}$ \\
$a_{35} [mm^{-7}]$ & $5.074361\times10^{-15}$ & $1.473937\times10^{-18}$ \\
$a_{37} [mm^{-7}]$ & $1.765092\times10^{-14}$ & $1.229728\times10^{-17}$ \\
$a_{39} [mm^{-7}]$ & $1.090339\times10^{-14}$ & $7.920244\times10^{-18}$ \\
$a_{41} [mm^{-7}]$ & $2.107488\times10^{-15}$ & $1.915300\times10^{-18}$ \\
$a_{43} [mm^{-7}]$ & $1.563836\times10^{-16}$ & $2.392989\times10^{-19}$ \\
$a_{45} [mm^{-8}]$ & $2.023527\times10^{-17}$ & $1.549090\times10^{-21}$ \\
$a_{47} [mm^{-8}]$ & $-1.013332\times10^{-16}$ & $-6.021327\times10^{-20}$ \\
$a_{49} [mm^{-8}]$ & $-4.288039\times10^{-17}$ & $-3.220436\times10^{-20}$ \\
$a_{51} [mm^{-8}]$ & $-4.547038\times10^{-17}$ & $-3.067110\times10^{-20}$ \\
$a_{53} [mm^{-8}]$ & $5.446197\times10^{-19}$ & $-1.145759\times10^{-21}$ \\
$a_{54} [mm^{-9}]$ & $-1.852556\times10^{-18}$ & $7.361501\times10^{-23}$ \\
$a_{56} [mm^{-9}]$ & $-8.888783\times10^{-18}$ & $-6.024621\times10^{-22}$ \\
$a_{58} [mm^{-9}]$ & $-1.033261\times10^{-17}$ & $-4.702521\times10^{-22}$ \\
$a_{60} [mm^{-9}]$ & $-3.380279\times10^{-18}$ & $8.548947\times10^{-23}$ \\
$a_{62} [mm^{-9}]$ & $-3.325001\times10^{-19}$ & $2.253766\times10^{-22}$ \\
$a_{64} [mm^{-9}]$ & $1.534227\times10^{-20}$ & $4.743691\times10^{-23}$ \\
\hline
\end{tabular}
\end{center}
\begin{tabnote}
\hskip18pt $^{\rm a}$ The zero value coefficients are not listed.\\
\end{tabnote}
\end{table}

Fig. 1 shows the optical layout of the telescope with appropriate parameters and coordinate systems. The optical axis ray (OAR) is the ray that passes through both the center of the aperture stop and the telescope focus \citep{chang03}. OAR plays a similar role to the optical axis of an on-axis system. The primary and the secondary mirror coordinate systems, which are denoted by $x_{p}-y_{p}-z_{p}$ and $x_{s}-y_{s}-z_{s}$, respectively, have their origins at the OAR's reflection points. Also, their z-axes are perpendicular to the mirror surfaces. The distance between the primary mirror and its focus along the OAR's extension is denoted by $l_{p}$. $l_{p}$ is negative since the focus is located behind the mirror. The secondary mirror has two foci: one is a common focus of the primary and the secondary mirrors and the other is the systems focus. The distances between the secondary mirror and the common (the system) foci along the OAR are denoted by $l_{s}$ and $l_{s}'$. $i_{1}$ and $i_{2}$ are the reflection angles of the OAR at the primary and the secondary mirrors, respectively. The system parameters of the telescope are listed in Table~1.

\begin{table}[t]
\begin{center}
\centering
\caption{Specifications of the Schwarzschild-Chang type wide-field off-axis telescope \label{tbl3}}
\doublerulesep2.0pt
\renewcommand\arraystretch{1.5}
\begin{tabular}{cc}
\hline \hline
Parameter      &      Value \\ \hline
Wavelength coverage & 8 - 12 \micron\\
Diameter of primary mirror & 70 mm\\
Diameter of secondary mirror &130 mm\\
Effective focal length &100 mm\\
Entrance pupil diameter &50 mm\\
Pixel size &45 \micron\\
Pixel FOV &92.8 \arcsec\\
Detector array format &320 $\times$ 240\\
Detector array FOV &8.2 $\arcdeg$ $\times$ 6.2 $\arcdeg$\\ \hline
\end{tabular}
\end{center}
\end{table}

Aperture stop is located at the secondary mirror to limit the secondary mirror size, which is already larger than the primary mirror (see Fig. 2). It can be shown that \citep{chang2006} the tangential and sagittal image planes are tilted in opposite directions from the median image plane by an angle $\theta_{t}$, which is given by
\begin{equation}
\theta_{t} = \nonumber \arctan[\frac{l_{s}'}{l_{s}}(\nonumber \tan i_{1}-\frac{l_{s} + l_{s}'}{l_{s}'}\nonumber \tan i_{2})] .
\end{equation}
In general, an off-axis two-mirror telescope has non-zero linear astigmatism due to this image plane tilt. However, it is possible to eliminate linear astigmatism by properly configuring the off-axis system. It is obvious from the above equation that the tilt angle of the tangential and sagittal image planes becomes zero when the system satisfies the following equation.
\begin{equation}
\nonumber \tan i_{1} = \frac{l_{s} + l_{s}'}{l_{s}'}\nonumber \tan i_{2}
\end{equation}
One can easily confirm from Table~1 that the designed telescope satisfies Eq. (2). Hence, the telescope has zero linear astigmatism. Furthermore, since the linear astigmatism is eliminated, the image plane is perpendicular to OAR \citep{chang05, chang2006}.

The surfaces of the primary and the secondary mirrors can be represented by xy polynomial in their coordinate systems as given in Eq. (3).
\begin{eqnarray}
\nonumber z(x,y)= a_{0}x + a_{1}y + a_{2}x^{2} + a_{3}xy + a_{4}y^{2} + a_{5}x^{3}
\end{eqnarray}
\begin{equation}
+ a_{6}x^{2}y + a_{7}xy^{2}+ a_{8}y^{3} + \cdots\cdots + a_{64}y^{10}
\end{equation}
The optimized polynomial coefficients of the primary and the secondary mirror surface equations are given in Table~2. The coefficients for the terms with odd powers of x are all zero since the system is symmetric to $y_{p}$-$z_{p}$ plane, which is also $y_{s}$-$z_{s}$ plane. Note that the coordinate system in Chang \& Prata (2005) and Chang (2006) is symmetric to y = 0 plane.

In our Schwarzschild-Chang telescope design, the diameter of the secondary mirror is much bigger than the primary mirror. Since the stop is located at the secondary mirror, the primary mirror is also larger than the entrance pupil diameter (EPD). Even though this design is not appropriate for a large aperture telescope, we can apply it to a small aperture telescope with a wide FOV. The EPD of this telescope is 50 mm and the focal ratio is 2. We plan to use our telescope with a 320 $\times\ 240$ microbolometer Infrared Focal Plan Array (SOFRADIR Inc.) with pixel size of 45 $\micron$. Then we will have a FOV of 8.2 $\arcdeg$ $\times$ 6.2 $\arcdeg$ by using only two reflective mirrors (see Table~3).

\begin{figure}[t]
\centering
 \epsfxsize=8cm
 \epsfbox{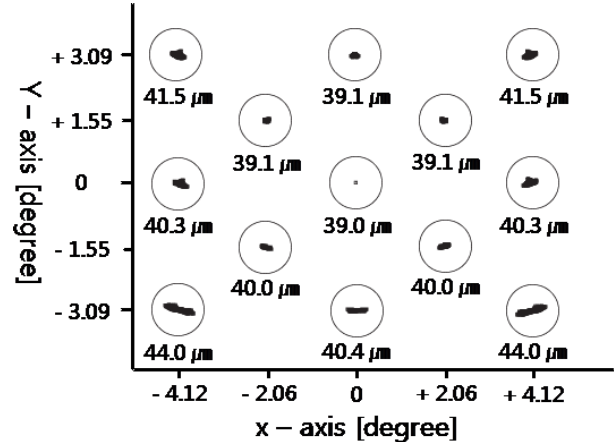}
\caption{Spot diagram of the Schwarzschild-Chang type off-axis telescope at various field positions. The diffraction effect is not considered in this diagram. The size of the circles represents that of the airy disk. The numbers under the circles are 80 \% EED including the diffraction effect.}
\end{figure}

Fig. 3 shows a spot diagram and encircled energy diameter (EED) simulated with a ray tracing software, Code V\footnote{Code V is the optical design and simulation software by Optical Research Associates.}. The spot diagram, not considering the diffraction effect, shows that most of the rays are enclosed within the detector pixel and the airy disk size. The airy disk size of the Schwarzschild-Chang type wide-field off-axis telescope is 49 $\micron$ at 10 $\micron$ wavelength. The EED includes the diffraction effect and 80 \% EED at each fields are minimum 39 $\micron$, maximum 44 $\micron$ (see Fig. 3). Modulate transfer function (MTF) results from Code V include the diffraction effect. MTF of our off-axis telescope is expected to be higher than 70 \% at the resolution limit of 11 cycle/mm, which is defined by the detector pixel size (see Fig.~4).

\begin{figure*}[p]
\centering \epsfxsize=12cm \epsfbox{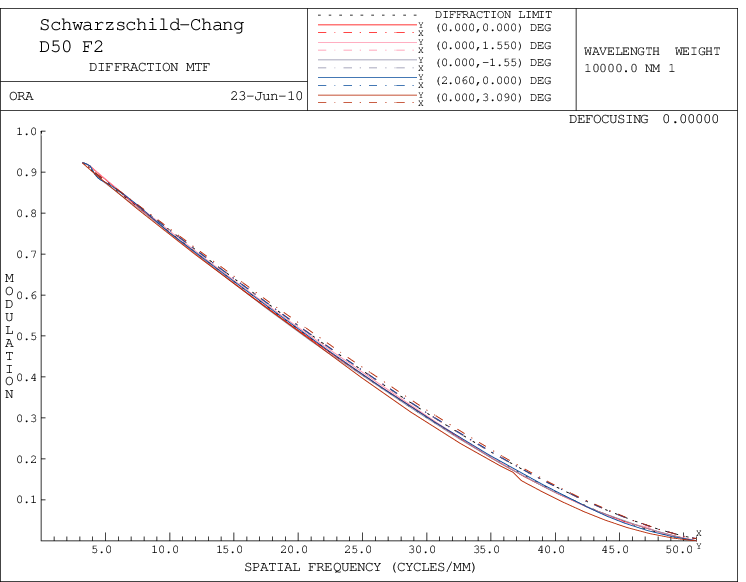}
\centering \epsfxsize=12cm \epsfbox{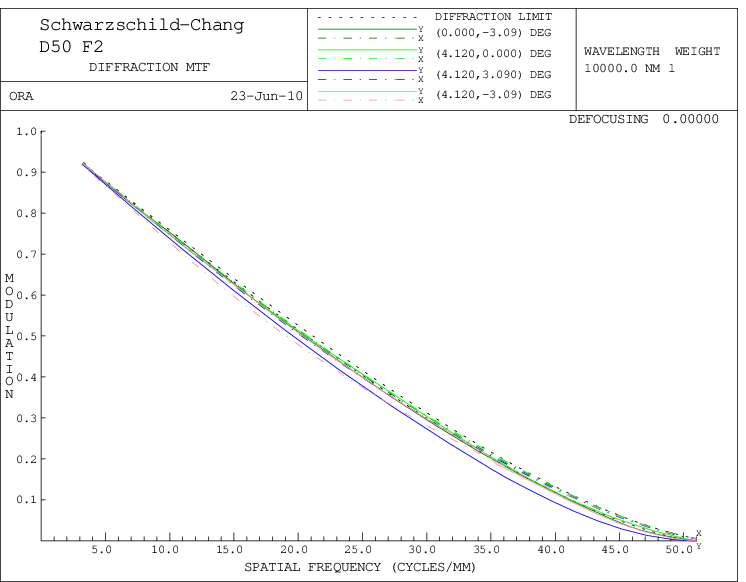}
\caption{MTF of the Schwarzschild-Chang type off-axis telescope at various field positions. The diffraction limit at 10 $\micron$ wavelength is shown. Pixel size of the detector is 45 $\micron$. We do not plot x-values smaller than 3cycle/mm.}
\end{figure*}

\section{FABRICATION OF OFF-AXIS MIRRORS}

\subsection{Machining of Off-Axis Surfaces}

The off-axis mirrors were made from aluminum (Al6061-T6). To minimize the machining time, we fabricate spherical mirrors first, which have differences of less than 0.2 mm from the designed surface shapes. Then, we fabricate the off-axis mirrors using a diamond turning machine (DTM), Freeform 700A (Precitech Inc.), which is a 5-axis control ultra precision DTM. We control three axes and aerostatic spindle to rotate a mirror, an x-axis linear servo floating on a hydrostatic bearing to laterally move the spindle together with the mirror, and a z-axis linear hydrostatic servo to translate the diamond tool for generating a depth profile \citep{kim07}. Rotation of the c-axis (spindle) is controlled in a low speed of about 50 RPM so that we can control the three axes to synchronize the spin of C-axis and position of the diamond bite (see Fig. 5). The paper discussing details of the mechanical structure of the mirrors and the fabrication processes is in preparation.

\subsection{Measurement of Mirror Surfaces}

Surface roughness of the fabricated off-axis mirrors (Fig. 6) was measured by an optical surface profilometer (NT 2000, Wyko Inc.). The central line averages of surface roughness, \textit{Ra}, of the primary and secondary mirrors are 4.21 nm and 4.24 nm, respectively, and the root mean square (RMS) values of surface roughness, \textit{Rq}, are 5.29 nm and 5.61 nm, respectively. Main impact of surface roughness is loss of light due to scattering. Total integrated scatter (TIS) is the fraction of scattered light and given by the following equation \citep{geary93, juergens05, ryu00}.
\begin{equation}
TIS = (\frac{4{\pi}\textit{Ra}}{\lambda})^2
\end{equation}
Thus we expect scattering loss on each fabricated mirror surface to be less than 1 \% at the IR wavelengths.

\begin{figure}[t]
\centering
\includegraphics[scale=1.00]{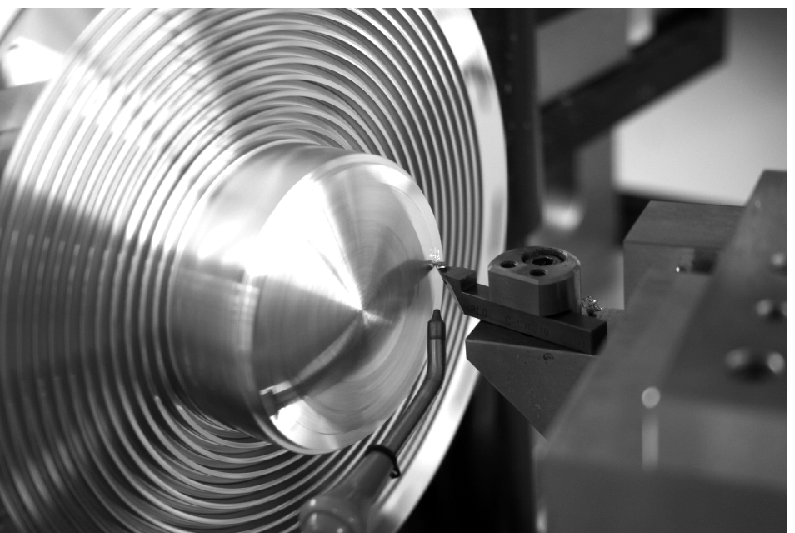}
\caption{Machining of the primary mirror using Freeform 700A. The nozzle seen under the diamond bite ejects mist (oil and air) to eliminate chips easily.}
\end{figure}

\begin{figure}[t]
\centering
\includegraphics[scale=1.00]{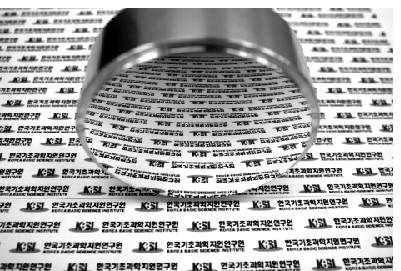}
\includegraphics[scale=1.00]{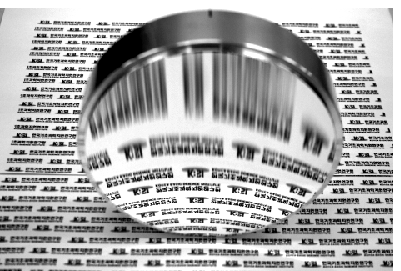}
\caption{Fabricated primary mirror (left) and secondary mirror (right) of the off-axis telescope.}
\end{figure}

Form accuracy of the off-axis mirrors was measured in peak-to-valley (P-V) and RMS values which are the differences between the designed surface function and the fabricated mirror shape. Since P-V values are easily influenced by vibration, scratch, or mote, RMS value are usually used to estimate quality of fabricated mirrors \citep{kim03}. The form accuracy was measured using an ultra-high accurate 3-D profilometer (UA3P, Panasonic Corp.) which uses an atomic force probe (AFP). UA3P makes AFP approach to the territory where repulsive atomic force is generated against the measurement surface and always keeps the atomic force constant. As a result, UA3P can measure the object without concerning about reflectance of the measurement surface \citep{tsutsumi05}.

The form accuracy of the primary mirror is 0.87 $\micron$ P-V and 0.19 $\micron$ RMS, and that of the secondary mirror is 3.1 $\micron$ P-V and 0.72 $\micron$ RMS. A total optical path difference (OPD) is defined as root sum square (RSS) of RMS errors from different mirrors \citep{juergens05} and is calculated to be 0.74 $\micron$ for our mirrors. Since a typically acceptable OPD should be less than 0.07$\lambda$, an OPD criterion for our optical system would be 0.70 $\micron$ at 10 $\micron$ wavelength \citep{born80, smith00}.

We use IDL\footnote{IDL is the data visuliztion \& analysis program which is a registered trademark of ITT Visual Information Solutions.} to fit a surface function of Eq. 3 to the fabricated mirror surface. We measure 108635 and 385318 point data from the primary and secondary mirrors. Figures 7 \& 8 compare the designed mirror shapes and the fitted functions. Although the fitted mirror surfaces are off by hundreds of nanometers from the designed shapes along the periphery, the central parts agree well with each other. We apply derived surface functions to the Schwarzschild-Chang off-axis telescope using Code V. When we use fitted function of the primary mirror, 80 \% EED at each fields are minimum 39 $\micron$, maximum 45 $\micron$ and 75 $\micron$, 79 $\micron$ for the secondary mirror and, if we apply both mirrors, 80 \% EED at each fields are minimum 86 $\micron$ and maximum 89 $\micron$.

\begin{figure*}[p]
\centering
\includegraphics[scale=0.3]{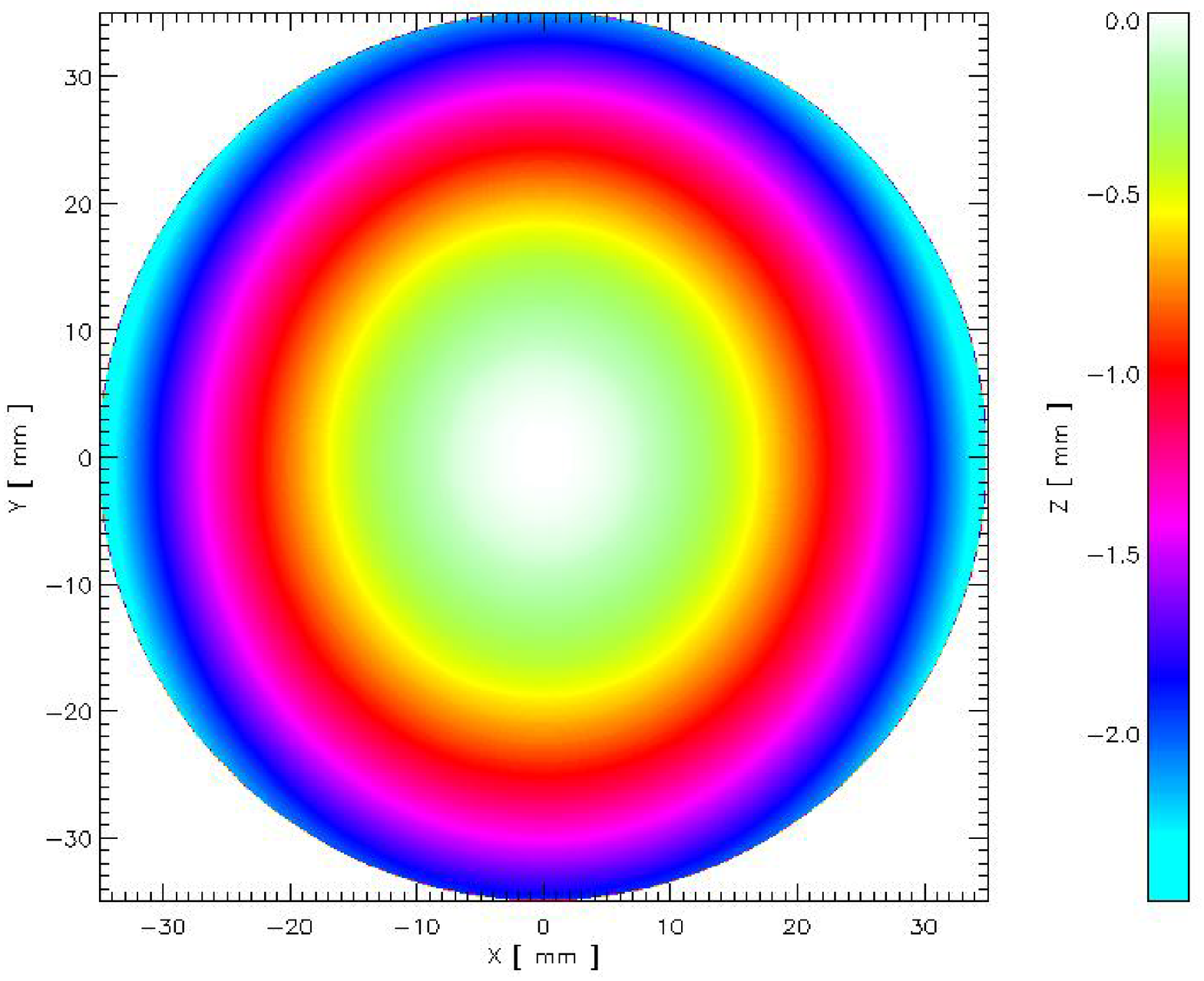}
\includegraphics[scale=0.3]{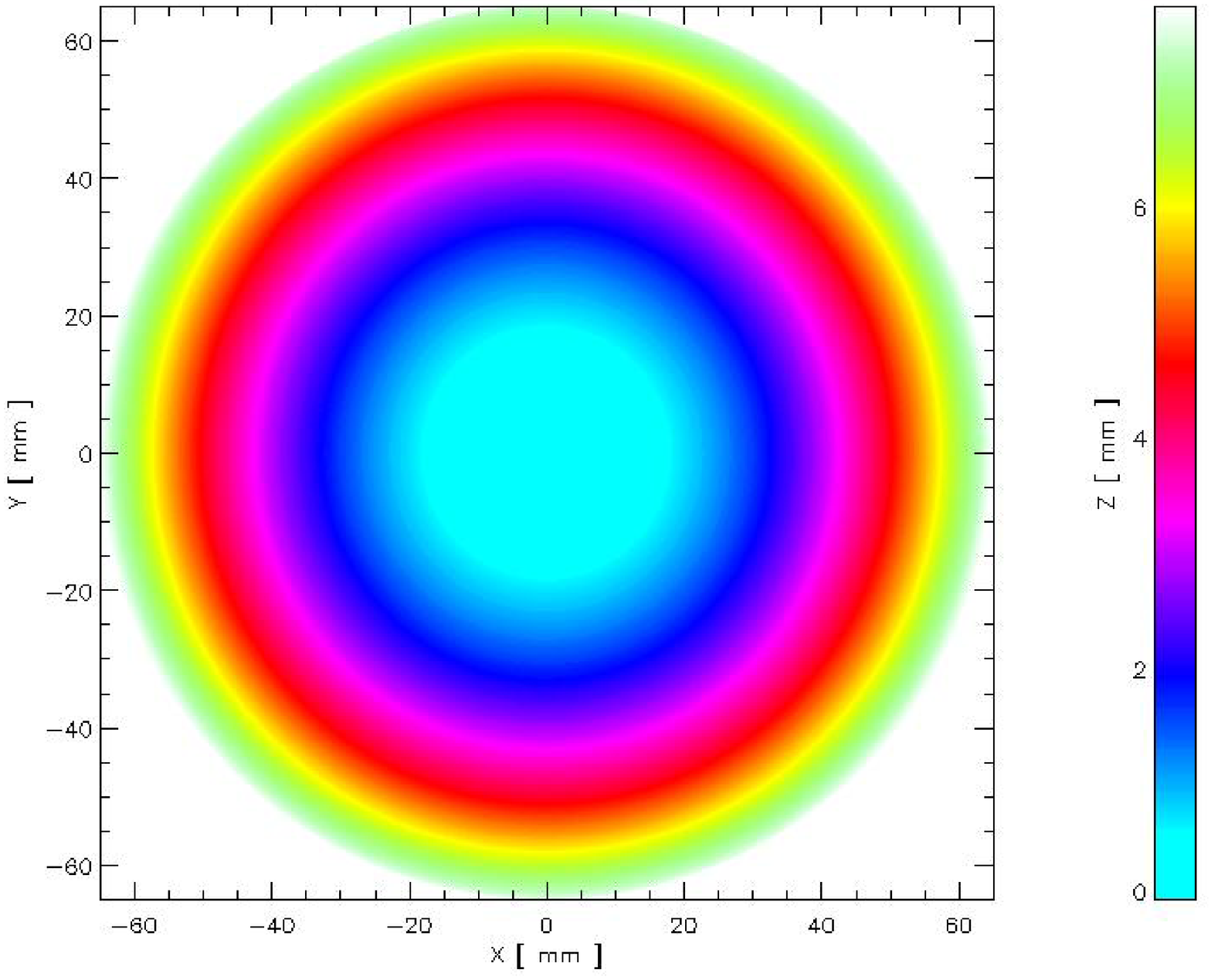}\\
\includegraphics[scale=0.3]{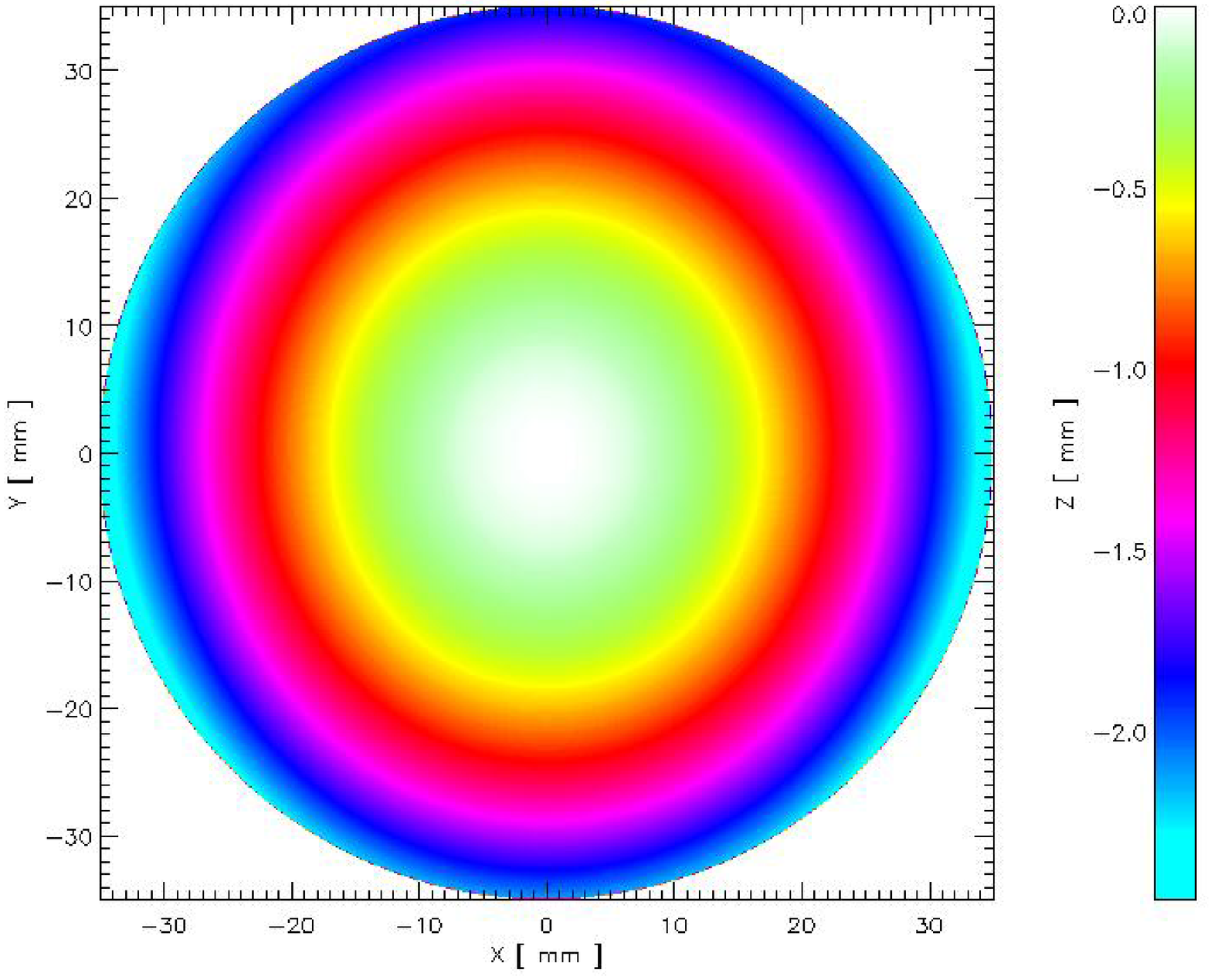}
\includegraphics[scale=0.3]{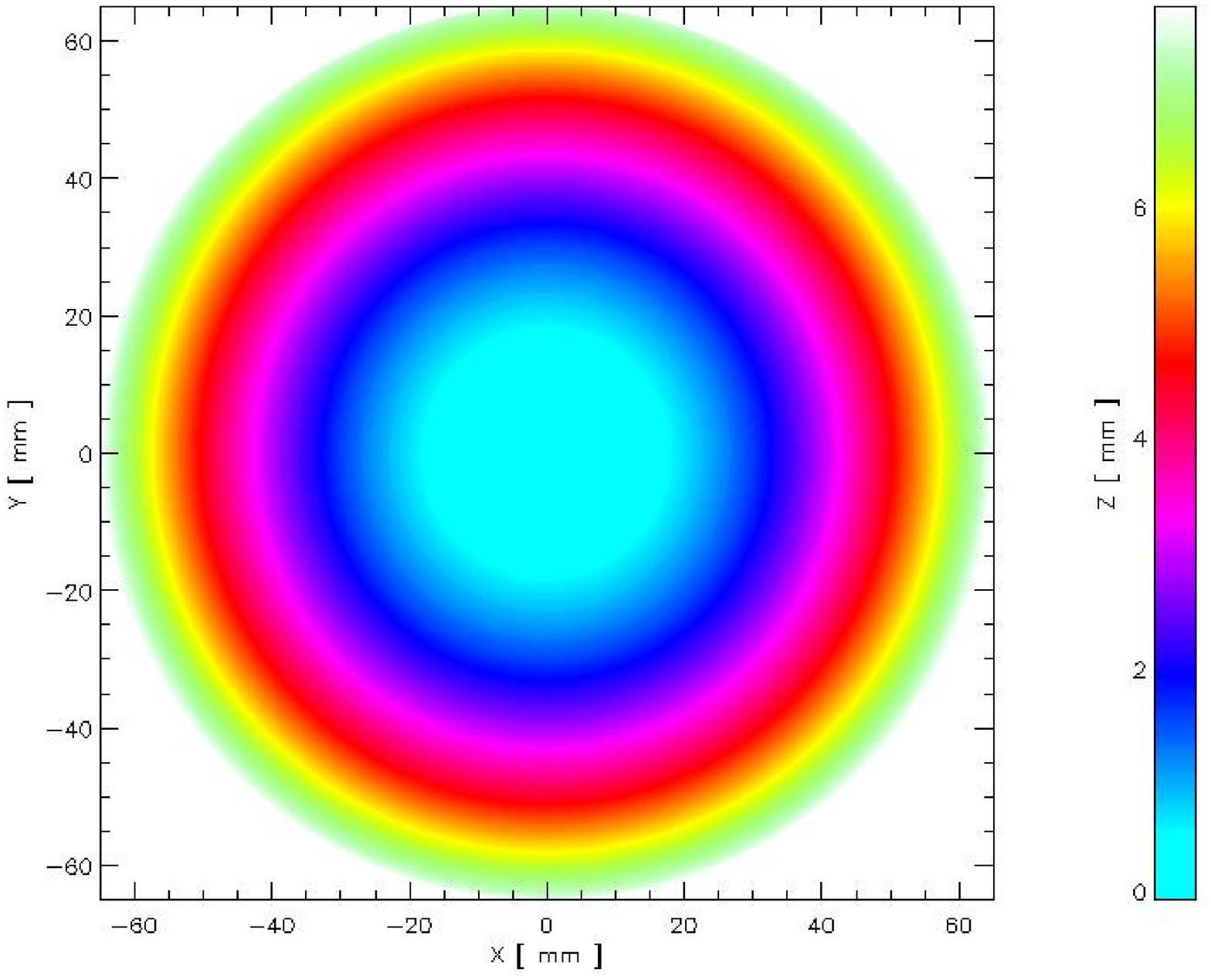}
\caption{Profile map of the mirror surfaces. Left images are for the primary mirror and the right images are for the secondary mirror. Images in the first and the second rows are for the designed and the fitted surface functions, respectively.}
\end{figure*}

\begin{figure*}[p]
\centering
\includegraphics[scale=0.3]{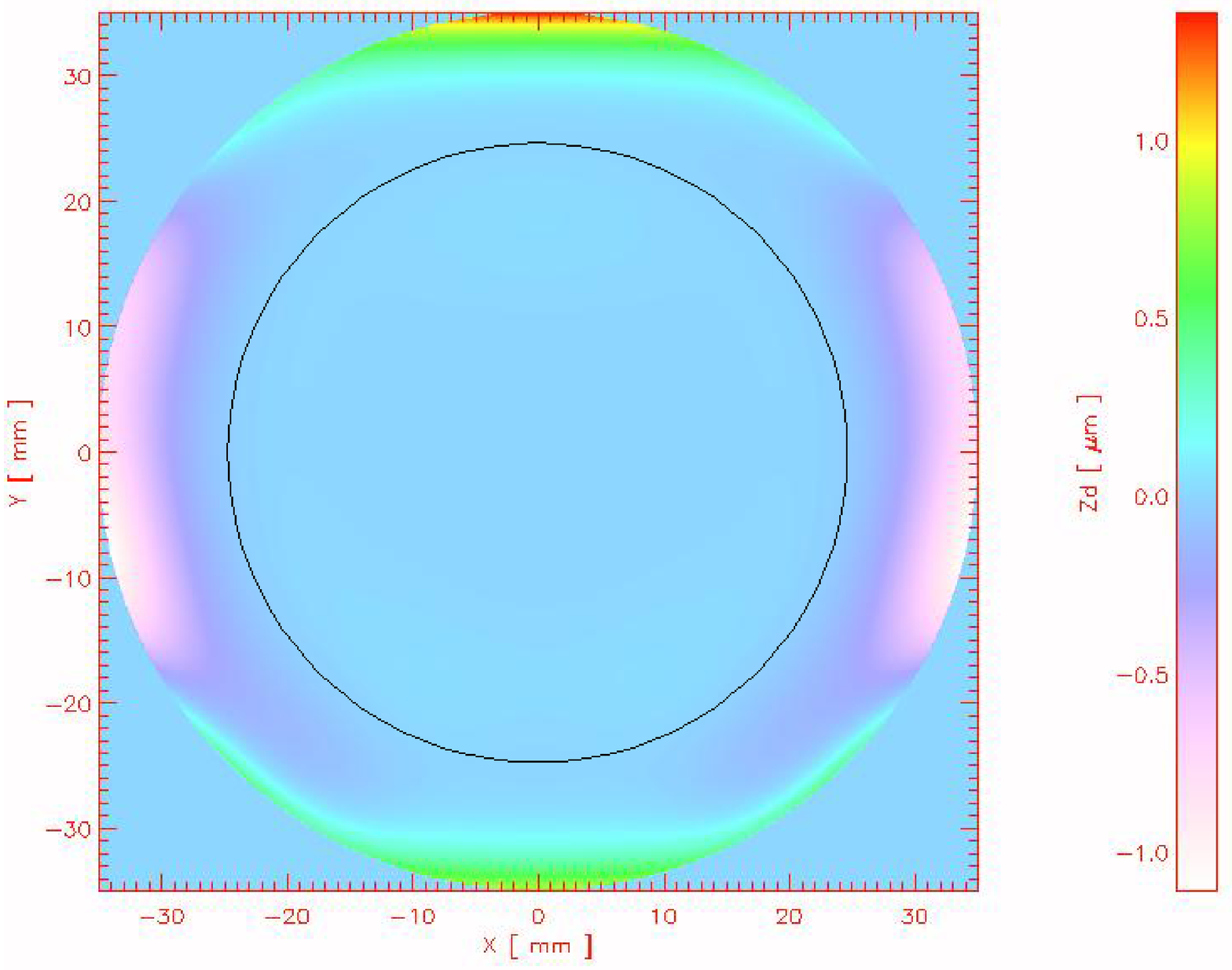}
\includegraphics[scale=0.3]{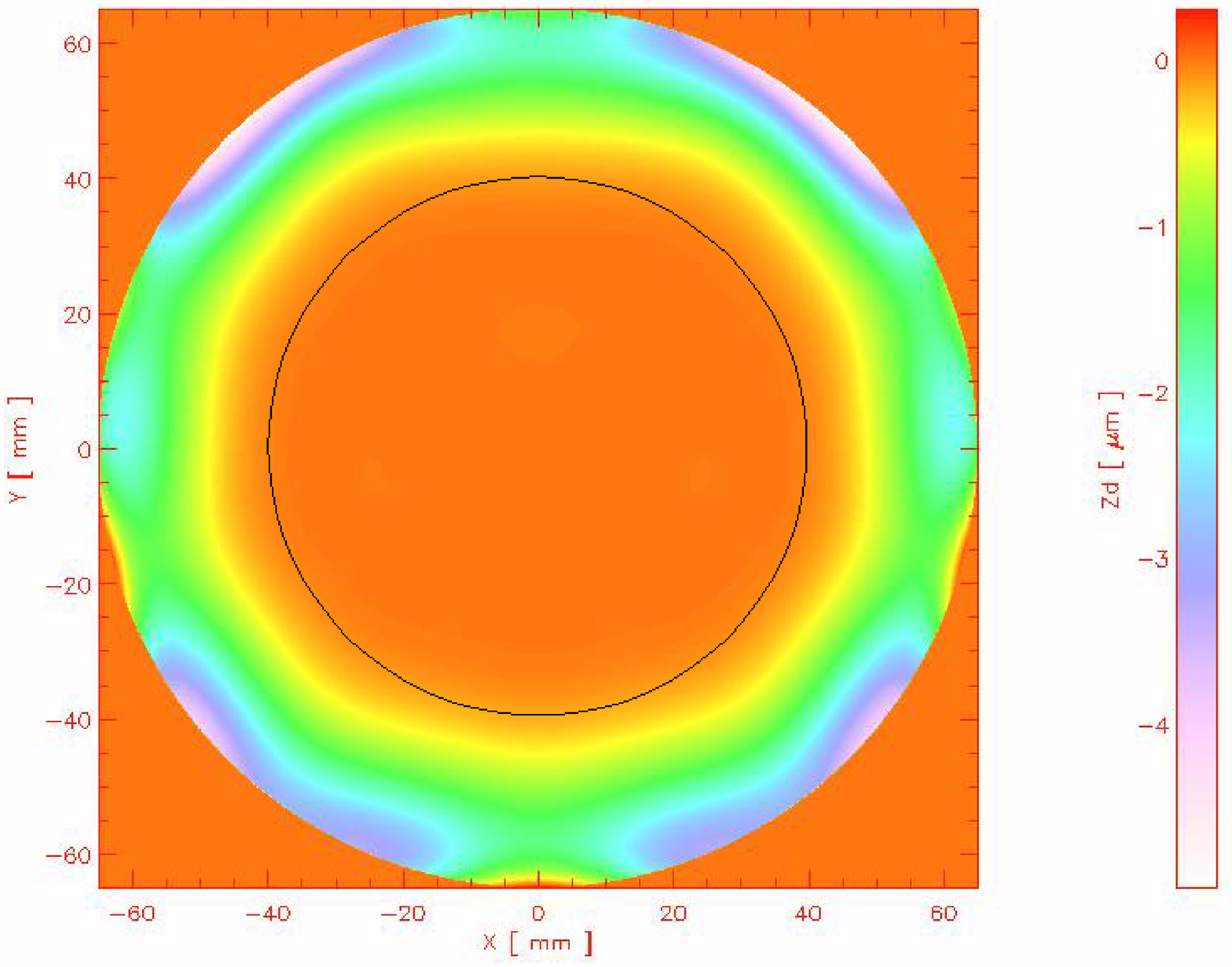}\\
\includegraphics[scale=0.3]{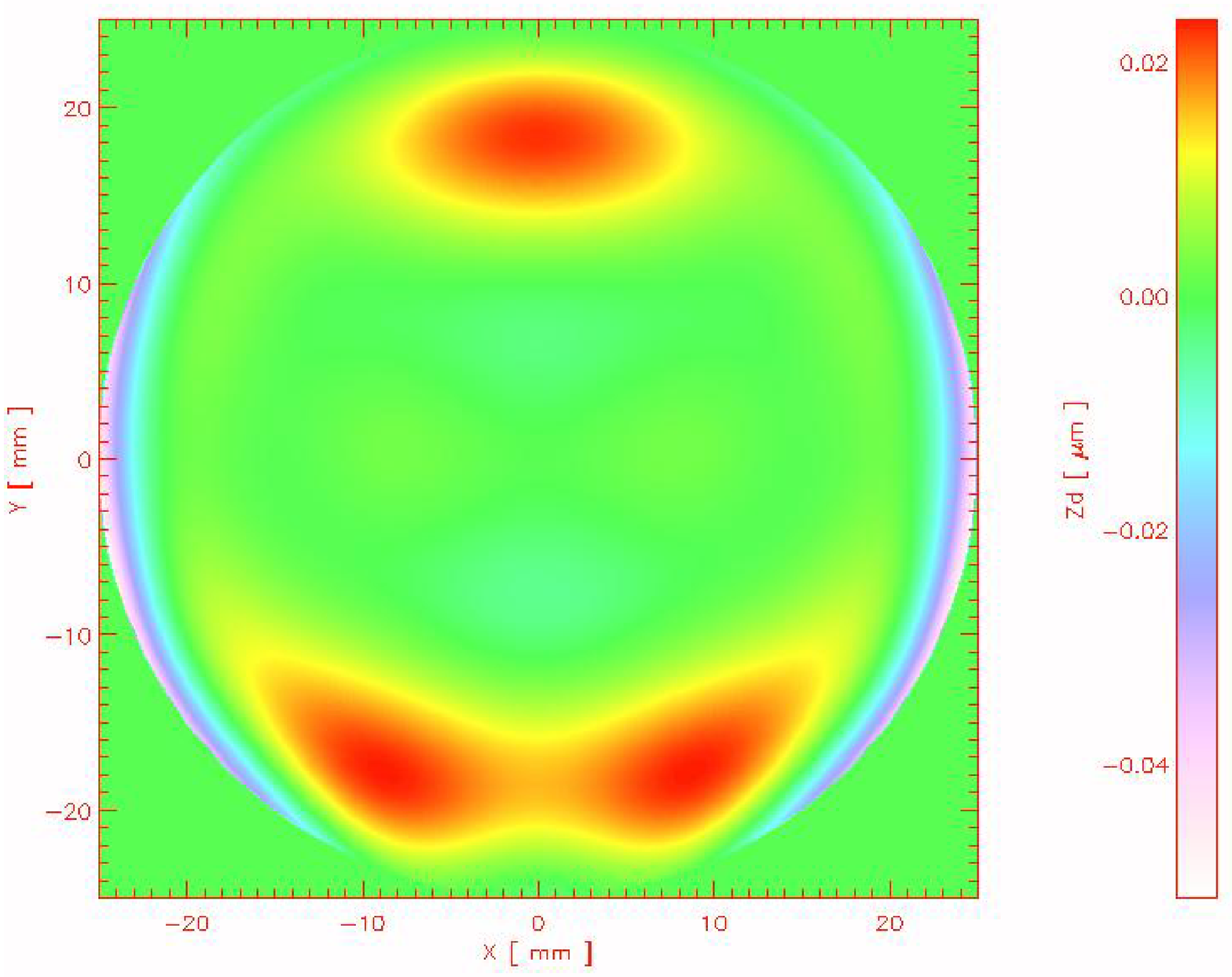}
\includegraphics[scale=0.3]{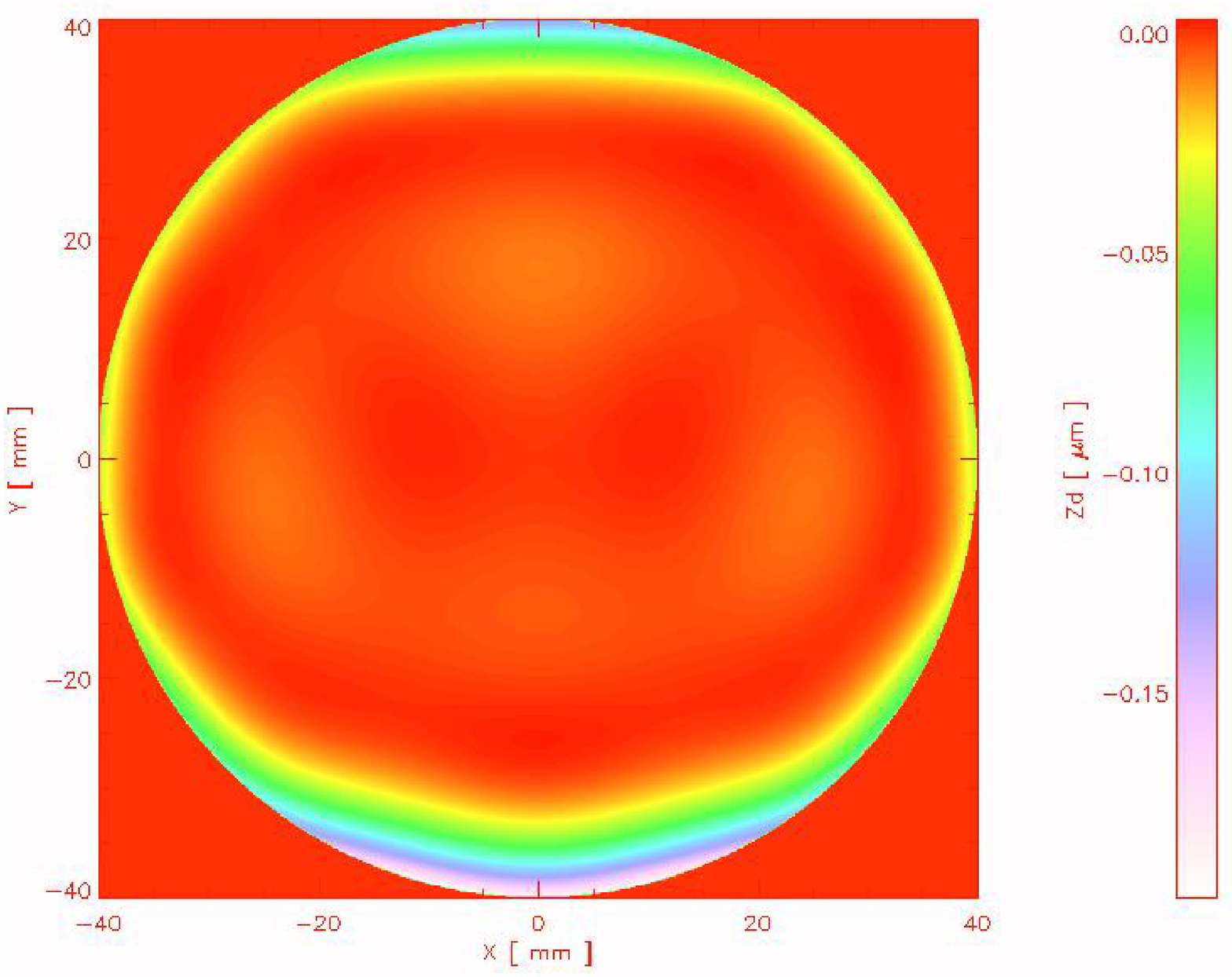}
\caption{Sag differences between the designed and the fitted surface functions for the primary mirror (left) and the secondary mirror (right). The lower images are zoomed in up of the central parts marked with the solid circles in the upper images.}
\end{figure*}

\subsection{Tolerance Analysis}

An off-axis system has more optical parameters than those of a typical on-axis reflecting system. Instead of a realistic Monte-Carlo tolerance analysis, we have checked an individual sensitivity of each optical parameter as the first order approach to the analysis. Since the encircled energy analysis by Code V takes into account both aberrations and the diffraction effect, we use the 80 \% EED as a practical performance measure in our tolerance analysis. Assuming a Nyquist pixel sampling, we set 90 $\micron$ for the maximum allowable limit of the 80 \% EED. We compensate image quality using the back focal length (BFL) which is the distance along the z-axis between the secondary mirror and the image plane.

We first estimate the manufacturing tolerance from our surface measurement results in section 3.2. And the alignment sensitivities were carried out for despace, decenter and tilt (see Fig. 9). Despace is defined as a displacement along the z-axis of the secondary mirror toward or away from the primary mirror. Decenter is defined as an error in position of the secondary mirror along the x- or y-axis. We do not consider z-axis for decenter which is eliminated by the compensator. For both mirrors, $\alpha$-, $\beta$- and $\gamma$-tilts are defined as rotations around x-, y- and z-axes, respectively \citep{schroeder00}. Table~4 shows the sensitivity analysis results, assuming that only single sensitivity parameter is perturbed while others are fixed.

\begin{figure}[!t]
\centering
\includegraphics[scale=0.4]{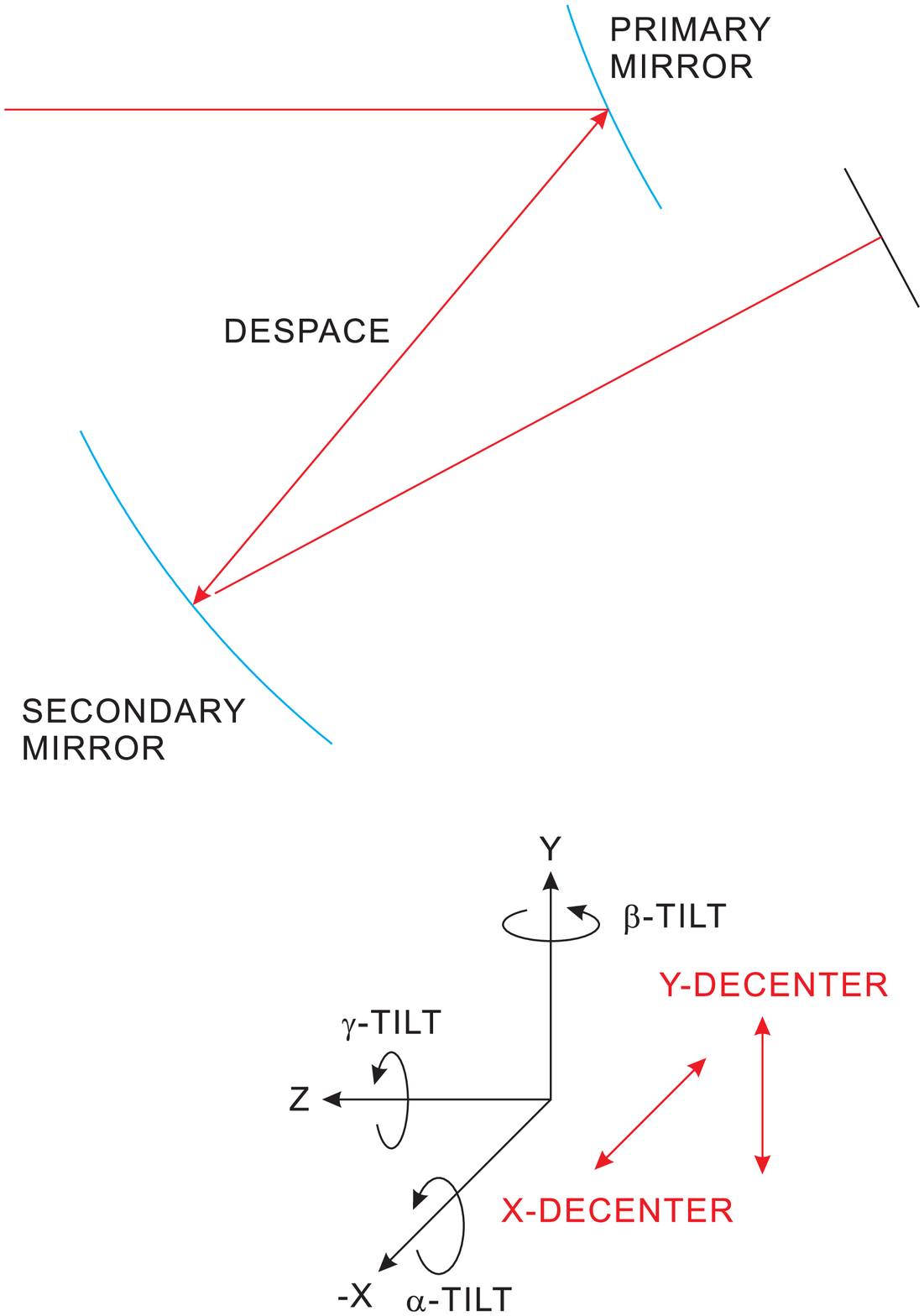}
\caption{Definition of tolerance parameters. Top image is for positional errors and bottom image for angular errors. Since the Code V coordinate system always assumes that the z-axis follows the optical axis, we define that the z-axis always aligns with the optical axis. Note that this x-y-z coordinate definition is different from $x_p-y_p-z_p$ and $x_s-y_s-z_s$ in Section 2. }
\end{figure}

\begin{deluxetable}{cccccc}
\tablecolumns{3}
\tablewidth{0pc}
\tablecaption{Sensitivity analysis results\label{tbl4}}
\tablehead{
\colhead{Tolerance Parameter} &
\colhead{Limit$^{\rm a}$} &
\colhead{EED$^{\rm b}$} &
\colhead{Compensation$^{\rm c}$} &
\colhead{Sensitivity$^{\rm d}$} &
\colhead{3$\sigma^{\rm e}$}
}
\startdata
Primary$^{\rm f}$ & 0.19 [$\micron$] & 45 [$\micron$] & 0.02 [mm] & 5.26 [$\micron/\micron$] & 2.43 [$\micron$]\\
Secondary$^{\rm f}$ & 0.72 [$\micron$] & 79 [$\micron$] & 0.12 [mm] & 48.61 [$\micron/\micron$] & 0.26 [$\micron$]\\
Despace & 3.55 [mm] & 90 [$\micron$] & 1.83 [mm] & 1.40 [mm/$\micron$] & 0.98 [mm]\\
x-axis decenter & 0.55 [mm] & 90 [$\micron$] & 0.00 [mm] & 83.64 [mm/$\micron$] & 0.15 [mm]\\
y-axis decenter & 0.55 [mm] & 90 [$\micron$] & 0.00 [mm] & 83.64 [mm/$\micron$] & 0.15 [mm]\\
Primary $\alpha$-tilt & 14.70 [$\arcmin$] & 90 [$\micron$] & 0.00 [mm] & 3.13 [$\micron/\arcmin$] & 4.07 [$\arcmin$]\\
Primary $\beta$-tilt & 15.90 [$\arcmin$] & 89 [$\micron$] & 0.00 [mm] & 2.83 [$\micron/\arcmin$] & 4.50 [$\arcmin$]\\
Primary $\gamma$-tilt & 37.80 [$\arcmin$] & 90 [$\micron$] & 0.00 [mm] & 1.22 [$\micron/\arcmin$] & 10.46 [$\arcmin$]\\
Secondary $\alpha$-tilt & 6.00 [$\arcmin$] & 90 [$\micron$] & 0.00 [mm] & 7.67 [$\micron/\arcmin$] & 1.66 [$\arcmin$]\\
Secondary $\beta$-tilt & 5.70 [$\arcmin$] & 87 [$\micron$] & 0.00 [mm] & 7.54 [$\micron/\arcmin$] & 1.69 [$\arcmin$]\\
Secondary $\gamma$-tilt & 38.10 [$\arcmin$] & 90 [$\micron$] & 0.00 [mm] & 1.21 [$\micron/\arcmin$] & 10.54 [$\arcmin$]\\
Detector $\alpha$-tilt & 72.00 [$\arcmin$] & 87 [$\micron$] & 0.03 [mm] & 7.67 [$\micron/\arcmin$] & 21.31 [$\arcmin$]\\
Detector $\beta$-tilt & 51.00 [$\arcmin$] & 90 [$\micron$] & 0.03 [mm] & 7.54 [$\micron/\arcmin$] & 14.11 [$\arcmin$]\\
\enddata
\tablenotetext{a}{Individual sensitivity limits for the 80 \% EED.}
\tablenotetext{b}{Size of the maximum 80 \% EED which is changed by tolerance (in ideal case, maximum 80 \% EED size is 44 $\micron$).}
\tablenotetext{c}{Compensator is the BFL.}
\tablenotetext{d}{Sensitivities for the 80 \% EED divided by the limit.}
\tablenotetext{e}{3$\sigma$ values corresponding to the allowed ranges to the tolerance parameters.}
\tablenotetext{f}{Form surface accuracy of mirrors in RMS.}
\end{deluxetable}

In order to derive the effective tolerance ranges, we use the quadrature sum of the individual sensitivity. Assuming that the tolerance merit, M, is a linear function of the parameters like
\begin{equation}
\nonumber M = \sum_{i=1}^Nc_{i}x_{i}
\end{equation}
where M is the merit, and $c_{i}$ is the sensitivity of a tolerance parameter $x_{i}$. We can also assume that each parameter is a random Gaussian variable with zero mean and non-zero variance. The variance of M is given by
\begin{equation}
\nonumber Var[M] = \sum_{i=1}^Nc_{i}^2Var[x_{i}]
\end{equation}
Considering a tolerance criteria (M $\pm$ dM), $\sqrt{Var[M]}$ is smaller than $dM/3$. Assuming that each tolerance parameter contributes equally to the effective tolerance ranges, we derive 3$\sigma$ values correspond to the minimum-maximum values of the allowed ranges of the tolerance parameters (see Table~4).

\section{ALIGNMENT OF THE OFF-AXIS TELESCOPE}

\subsection{Building of an Alignment System}

Off-axis telescopes are difficult to align because primary and secondary mirrors do not share the same optical axis and there is $\gamma$-tilt which does not exist in on-axis telescopes. Our optical design, however, has advantage that the OAR is perpendicular to the focal plane. To align the Schwarzschild-Chang type off-axis telescope, we use a laser diode (632.8 nm) for a ray tracing method. All parallel incident rays which trace the optical path of the telescope are predicted to be focused on a single point regardless of the incidence positions of the rays.

The alignment was conducted on an optical table installed with alignment tools such as a stop with 5 holes, a primary mirror and a secondary mirror blocks with central holes, and a flat mirror (see Fig. 10). The mirror blocks, made from 5 mm thickness plates, have central 1 mm holes for the incident rays. For the reflected ray, the primary block has an off-center 20 mm hole. The reflected ray on the secondary mirror is off from the block plate. These blocks are fabricated using a machining center with machining errors less than 10 $\micron$. These blocks are attached at 22.54 mm and 15.55 mm in front of the primary and secondary mirror surfaces respectively.

The initial calibration of the stage position is very important. So we first calibrate the gage of the stages using a height gage (Tesa micro-hite 600, TESA Technology Ltd.). And we have made very accurate mount base blocks. The machining errors are less than 10 $\micron$. The mount base blocks are assembled between mirror mounts and optical table. Using the mount base blocks, mirror mount stages are assembled accurately at the optical table.

After calibrating stage and their positions, the two telescope mirrors and the flat mirror at the focal plane are assembled on the mounting stages. For each mirror and mount, we used four pins to minimize assembly errors of tilt and decenter (see Fig. 11). The stages for primary, secondary and flat mirrors can adjust x-, y- and z-axis positions. Furthermore primary and secondary mirror mounts can change $\alpha$- and $\beta$-tilt of the mirror. We made a parallel incident ray reach the center of the primary mirror by adjusting tilt and translation of the laser. We set two pin holes in 600 mm range. The parallelism of the incident ray is given by
\begin{equation}
\triangle\Theta \approx \frac{\triangle x}{L_{1}} \approx \frac{0.3 mm}{600 mm} \approx 0.0005 rad  \approx 1.7 \arcmin
\end{equation}
where $\triangle$x is a residual error of the primary mirror and $L_{1}$ is the distance from the stop to the mirror block of the primary mirror (see Fig. 10). Since the size of the hole which is located in front of primary mirror is 1 mm, we expect that the pointing accuracy by eyes would be better than $1/3$ of the hole size. The assumption of 0.3 mm would be acceptable within a factor of 2, and the assumed values are mostly smaller than the tolerance values in Table~4. Thus, the parallelism of the incident ray is guaranteed less than $1.7\arcmin$ when $\triangle$x is less than 0.3 mm. Since the ray from the laser diode is not an ideal point source, (about 2 mm Gaussian shape), we set a pin hole with 1 mm in diameter in front of the laser source. Since the distance between the secondary mirror black and the focal plane is 240 mm, the divergence of the laser image on the flat mirror would be insignificant.

\begin{figure}[!h]
\centering
\includegraphics[scale=1.00]{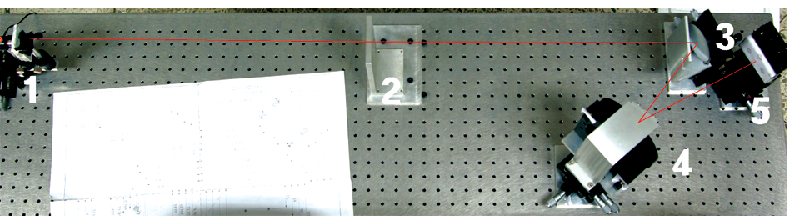}\\
\includegraphics[scale=0.30]{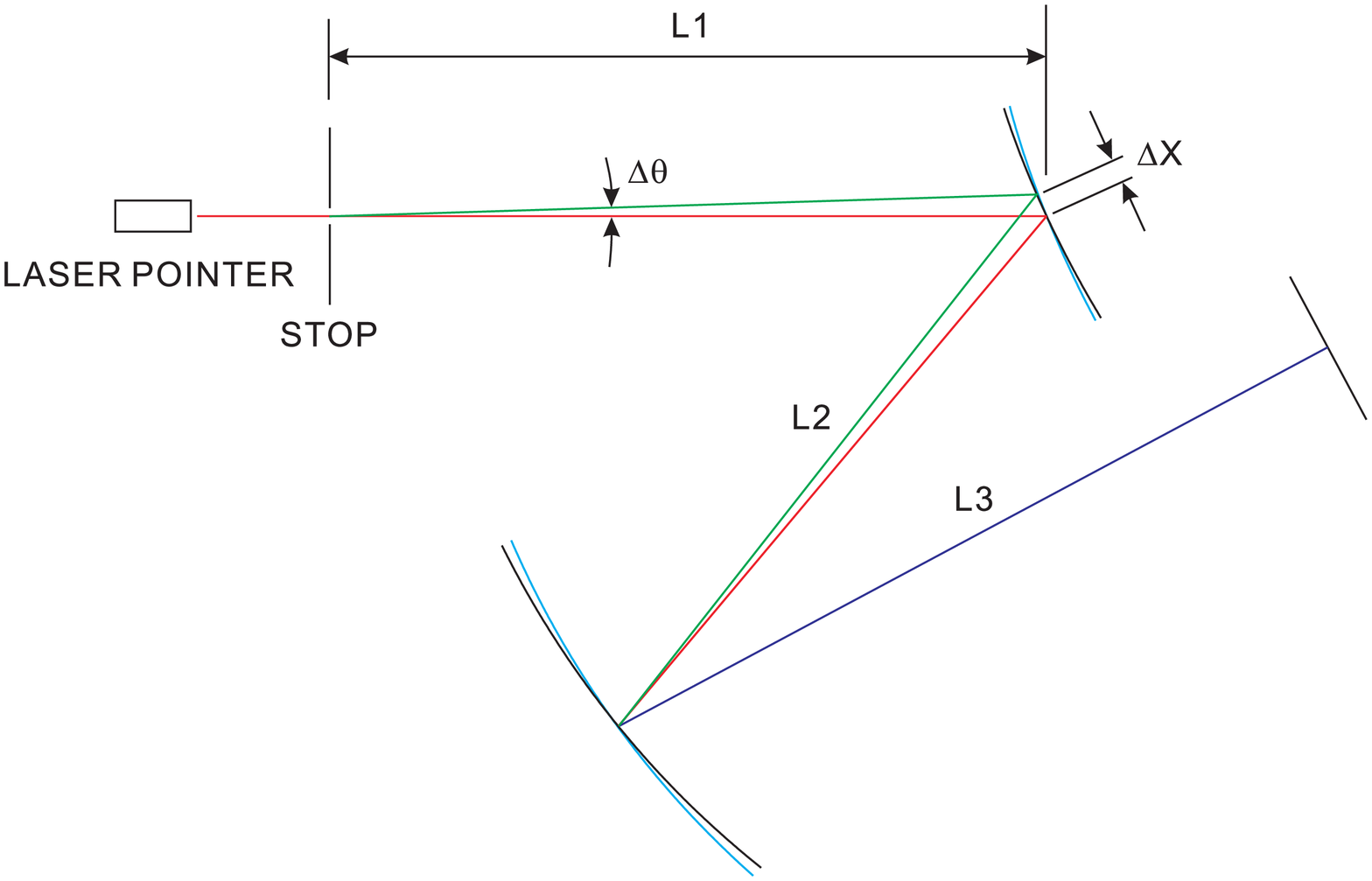}
\caption{Alignment system using a ray tracing method. Top image is a picture of the system; (1) the light source (laser diode), (2) the stop, (3) the primary mirror, (4) the secondary mirror, and (5) the flat mirror at the focal plane. The stop is fixed on the optical table. The mirror blocks are attached to the primary and secondary mirror stages and move together with mirrors. Bottom image is an illustration describing alignment accuracy expressed by Eq. (7).}
\end{figure}

\begin{figure}[!h]
\centering
\includegraphics[scale=1.00]{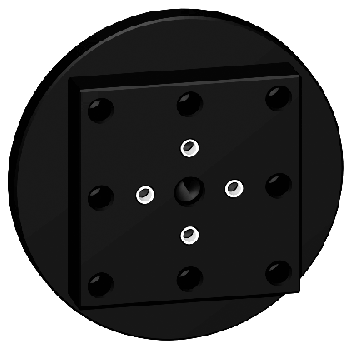}
\includegraphics[scale=1.00]{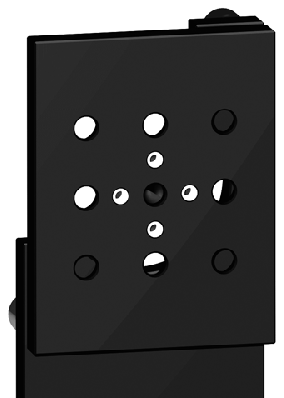}
\caption{Pinholes for alignment of mirror mounts. White holes are pinholes on the backside of the primary mirror(left) and a mounting stage(right).}
\end{figure}

\subsection{Ray Tracing Method for Alignment}

\begin{figure*}[p]
\centering \epsfxsize=14cm \epsfbox{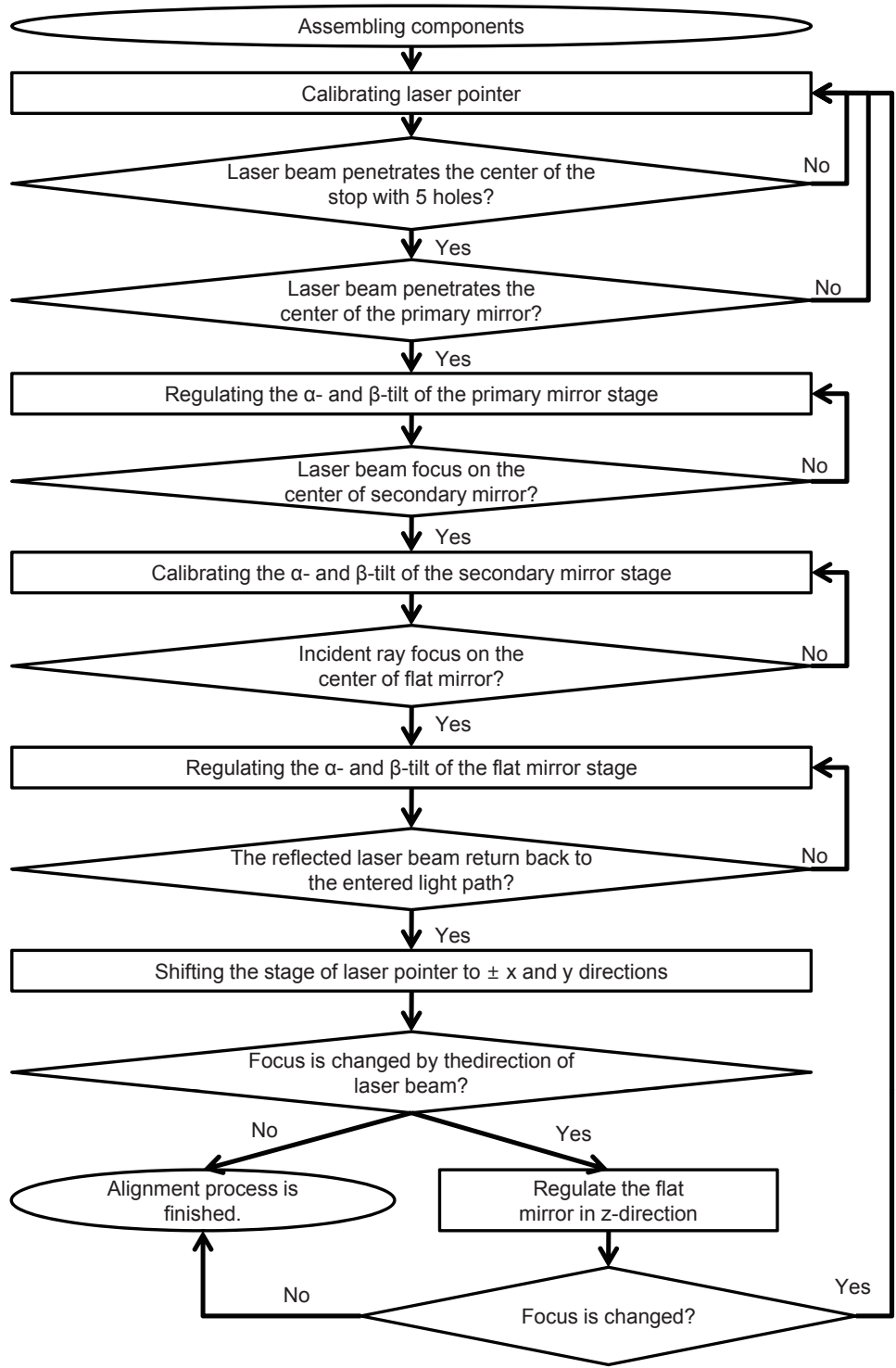}
\caption{Alignment process for the off-axis telescope using a ray tracing method.}
\end{figure*}

The flowchart in Fig. 12 shows the ray tracing method we use for the alignment. As has been described in Section 4.1, we first adjust x, y, and z positions of the mirror mounting stages and regulate their $\alpha$ and $\beta$ tilting angles. Then we remove the stop and the primary and the secondary mirror blocks. After that, we shift position of the laser to $\pm$ x and y directions and make sure that the focus does not move (see Fig. 13). As shown in Fig. 14, we use a stop with 5 holes to increase the optical path length (OPL) further, since the accuracy of a ray tracing method is directly proportional to the OPL. The incident ray through the positive y hole on the stop is reflected by the flat mirror at the focal plane and is returned back to the negative y hole at the stop. This ensures that the system is properly aligned, since the system is designed to have zero linear astigmatism and an off-axis optical system of which the linear astigmatism is eliminated has an axially symmetric property around the optical axis ray \citep{chang05}.

\begin{figure}[t]
\centering
\includegraphics[scale=1.00]{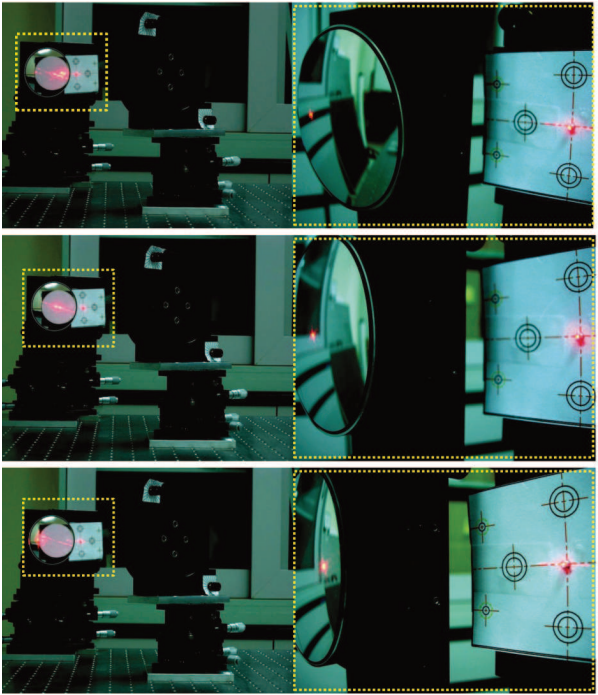}
\caption{Final stage of the alignment process. Position of the focus does not move regardless of different positions of incident rays on the primary mirror; a negative y position (top), the center (middle), and a positive y position (bottom).}
\end{figure}

\begin{figure}[t]
\centering
\includegraphics[scale=0.95]{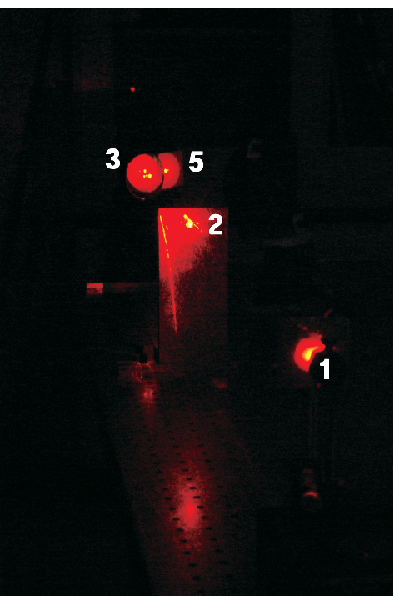}
\includegraphics[scale=0.95]{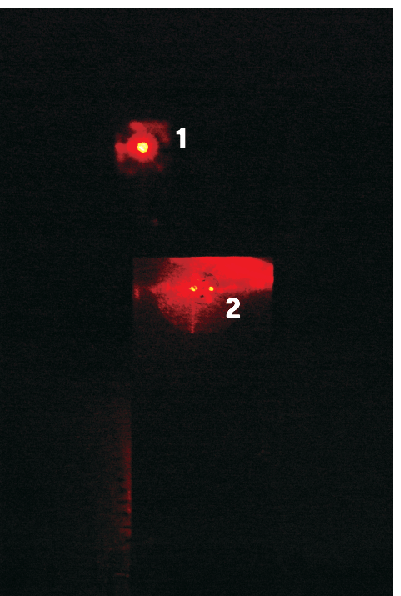}
\caption{Confirmation of the alignment using the stop with 5 holes. Left image shows the laser coming out through the negative y hole of the stop and returning back to its positive y hole; (1) the light source (laser diode), (2) the stop with 5 holes, (3) the primary mirror, (4) the secondary mirror, which is too dark to be seen in this picture, and (5) the flat mirror at the focal plane. Right image is a zoom-in of the back side of the stop. We can see that the negative y hole (left) and the positive y hole (right) on the stop are illuminated by the laser.}
\end{figure}

\section{POINT IMAGE TEST}

We made a point object using a laser diode and an optical jumper cord. Wavelength of the laser is 632.8 nm. We use the connecting tip of a typical optical jumper cord as the point source. The diameter of the core is 7 $\micron$ and numerical aperture is about 0.1. The source is located 1250 mm in front of the primary mirror.
The flat mirror at the focal plane is replaced with a testing detector, a 3K $\times$ 2K CCD with 3.5 $\micron$ pixels (Lumenera Corp.).

We performed a Code V simulation by applying the same testing condition to the optical design. The surface functions derived from the fabricated mirrors (Fig.~7) were used to modify the Code V design to be more realistic. Predicted 80 \% EED of the image of a point object is 86 $\micron$ and its RMS size is 119 $\micron$ (see Fig. 15). Compared to this, measured size of the real point source image (Fig. 16) generated by our manufactured telescope is about 100 $\micron$. We conclude that the prediction and the testing result are consistent with each other in order of magnitudes.

\begin{figure}[t]
\centering \epsfxsize=8cm \epsfbox{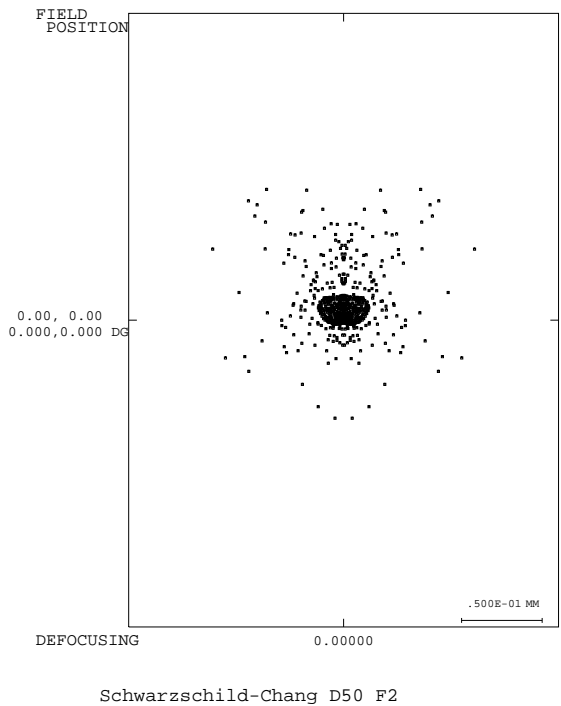}
\caption{Simulated image for the point image test. The optical model of the off-axis telescope was modified by using the fitted surface functions in Fig. 7. RMS size of the spot is 119 $\micron$.}
\end{figure}

\begin{figure}[t]
\centering \epsfxsize=8cm \epsfbox{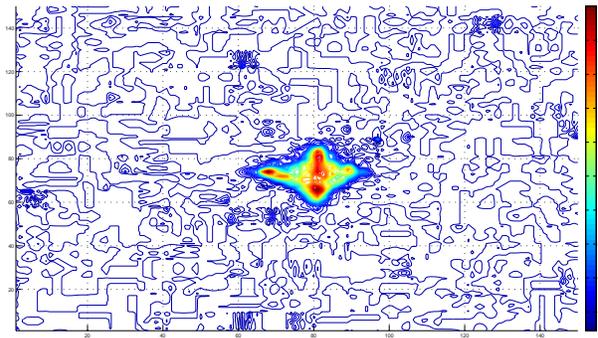}
\caption{Point image of the Schwarzschild-Chang type off-axis telescope.}
\end{figure}

\section{SUMMARY}{}

We have developed an off-axis reflective telescope for wide-field IR observations. The process and results of this project are summarized as follows.

1. The optical design is based on the Schwarzschild-Chang type telescope. Since the diameter of the secondary mirror is much larger than the EPD, this design is not appropriate for a large aperture telescope. Instead we can apply it to a small aperture telescope with a wide FOV. The EPD of the telescope is 50 mm and the FOV is 8.2 $\arcdeg$ $\times$ 6.2 $\arcdeg$ with the focal ratio of 2. We have performed tolerance analysis using EED as a performance measure and picked out sensitive parts in the optical alignment.

2. The off-axis mirrors were fabricated directly from aluminum plates using DTM at Korea Basic Science Institute (KBSI). Surface roughness of the mirrors is about 4 nm (\textit{Ra}). Form accuracy of the primary and secondary mirrors is 0.19  $\micron$ and 0.72  $\micron$ in RMS, respectively.

3. The measured data of the fabricated mirror surfaces were analyzed to fit surface functions. The derived surface functions were compared with the optical model.

4. We have accomplished alignment of the two-mirror off-axis telescope using a ray tracing method. The accuracy of our alignment system is about $1.7\arcmin$ and it is small enough compared to the alignment tolerances.

5. We have done a point image test using the aligned system. The quality of the obtained image is consistent with the prediction by the simulation.

\section{ACKNOWLEDGMENT}{}
This research was supported by WCU (World Class University) program through the National Research Foundation of Korea funded by the Ministry of Education, Science and Technology  (R31-10016).
S. Kim, G. H. Kim, S. C. Yang, and M. S. Kim were supported by Development of Ultra-Precision Thermal Imaging Microscope Project of KBSI.

\end{document}